\documentclass[12pt]{revtex4}
\usepackage{graphicx}
\usepackage{amsfonts}
\usepackage{array}

\def\DP{{\scriptscriptstyle\rm DP}}
\def\MF{{\scriptscriptstyle\rm MF}}

\def\xvec{{\bf x}}

\begin{document}
\title[Contact processes with long-range interactions]
      {Contact processes with long-range interactions}
\author{F. Ginelli$^1$, H. Hinrichsen$^2$, R. Livi$^{3,4,5}$, D. Mukamel$^6$,\\
and A. Torcini$^{4,5,7}$}

\address{$^1$ Service de Physique de l'Etat Condens{\'e}, CEA/Saclay, 91191 Gif-Sur-Yvette, France}
\address{$^2$ Fakult\"at f\"ur Physik und Astronomie, Universit\"at W\"urzburg, \\
    Am Hubland, 97074 W\"urzburg, Germany}
\address{$^3$ Dipartimento di Fisica, Universit\`a di Firenze, Unit\'a INFM \\ via Sansone 1 - I-50019 Sesto Fiorentino, Italy}
\address{$^4$ Centro Interdipartimentale per lo Studio delle Dinamiche Complesse \\ via Sansone 1 - I-50019 Sesto Fiorentino, Italy}
\address{$^5$ Sezione INFN di Firenze\\ via Sansone 1 I-50019 Sesto Fiorentino, Italy}
\address{$^6$ Department of Physics of Complex Systems, Weizmann Institute of Science,
Rehovot 76100, Israel}
\address{$^7$ Istituto dei Sistemi Complessi, CNR, \\ via Madonna del Piano 10, I-50019 Sesto Fiorentino, Italy}

\begin{abstract}
A class of non-local contact processes is introduced and studied
using mean-field approximation and numerical simulations. In these
processes particles are created at a rate which decays
algebraically with the distance from the nearest particle. It is
found that the transition into the absorbing state is continuous
and is characterized by continuously varying critical exponents.
This model differs from the previously studied non-local directed
percolation model, where particles are created by unrestricted
Levy flights. It is motivated by recent studies of non-equilibrium
wetting indicating that this type of non-local processes play a
role in the unbinding transition. Other non-local processes which
have been suggested to exist within the context of wetting are
considered as well.

\end{abstract}

\parskip 2mm
\vglue 5mm

\maketitle

\section{Introduction}

The contact process (CP) is known as a simple model for epidemic
spreading that mimics the interplay of local infections and
recovery of individuals~\cite{Mollison77,Liggett85}. It is defined
on a $d$-dimensional lattice whose sites could be either active
(infected) or inactive (non-infected), denoted as '1' and '0',
respectively. The model evolves random-sequentially by two
competing processes, namely, nearest-neighbor infections $01/10\to
11$, and spontaneous recovery $1 \to 0$. Depending on the relative
frequency of these moves the contact process displays a continuous
phase transition from a fluctuating active state into an absorbing
state which belongs to the universality class of directed
percolation
(DP)~\cite{MarroDickman,Hinrichsen00,OdorReview,Lubeck}.

In the present paper we investigate a generalized version of the
one-dimensional contact process where inactive sites can be
activated over long distances. We consider a lattice model where
sites could be either active or non-active. The model evolves by
random sequential updating with the following transition rates:
\begin{eqnarray}
\label{process1}
1  &\rightarrow& 0      \hspace{8mm}\mbox{with rate }\hspace{0.5cm} 1 \\
\label{process3}
0 &\rightarrow& 1   \hspace{7mm}\mbox{with rate }\hspace{0.5cm} q/l^\alpha
\end{eqnarray}
The first move corresponds to the usual local annihilation
process, while the second move describes a long range process in
which an inactive site which is located at a distance $l$ from the
nearest active one becomes active. The rate at which such a
process takes place decays algebraically with the distance $l$
from the nearest active site (measured in lattice site units),
reflecting the long range nature of the interaction. The overall
rate of this process is governed by the control parameter $q$,
while the characteristic shape of the interaction is controlled by
the exponent $\alpha$. In particular, usual short range dynamics
is recovered in the limit $\alpha \to \infty$. In what follows we
refer to this model as the $\alpha$-process. Note that its
definition is valid in any dimension although we are primarily
interested in the one-dimensional case.

\begin{figure}
\centerline{\includegraphics[width=114mm]{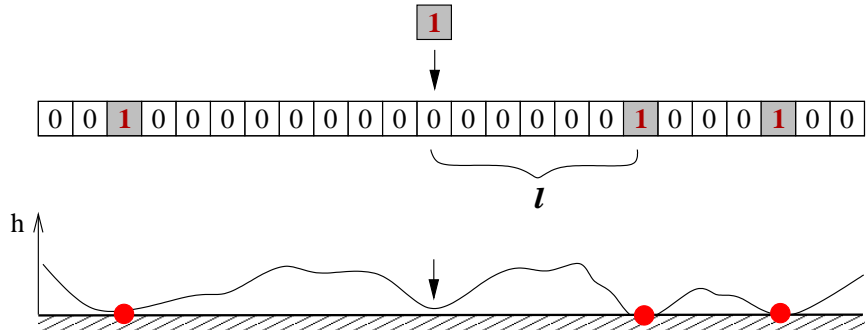}}
\caption{\label{fig:interface} \small
Long-range infections in the generalized one-dimensional contact process. Any site between two active sites can be activated with a rate proportional to $l^{-\alpha}$, where $l$ is the distance to the nearest active site. The model is motivated by the dynamics of a fluctuating interface growing on an inert substrate (see text).
}
\end{figure}

The long-range contact process investigated here is motivated by
recent studies of depinning transitions in non-equilibrium wetting
processes~\cite{Wetting1,GiadaMarsili,Candia,Munoz,Wetting2},
where one considers a fluctuating interface next to a hard-core
wall. Regarding the pinned domains of the interface as active
sites and unpinned domains as inactive ones, the dynamics of the
fluctuating interface may be projected onto that of a contact
process (see Fig.~\ref{fig:interface}). For example, unpinning of
the interface due to deposition would correspond to spontaneous
recovery $1\to 0$, while lateral growth of a pinned domain would
correspond to local spreading of activity by $01/10 \to 11$.
However, as unpinned regions of the interface fluctuate in the
bulk they may spontaneously become pinned to the substrate far
away from other binding sites. As illustrated in the figure, such
a pinning effectively leads to long-range infections in the
corresponding contact process. Clearly, the rate for pinning will
decrease as we move away from the binding site. Numerical studies
of some wetting models (not shown in this work) indicate that this
rate decays algebraically as $l^{-\alpha}$ with an exponent
$\alpha\approx 2.6$. This motivates the postulated power law in
Eq.~(\ref{process3}). Nevertheless the relation between
non-equilibrium wetting and the $\alpha$-process should be
considered as an analogy, rather than an exact mapping, because
the projection neglects the internal dynamics of unpinned regions
of the interface.

In recent years the contact process has also been generalized in
various other ways to include long-range
infections~\cite{MarquesFerreira94,Albano96}. In these studies it
is assumed that a given active site can activate \textit{any}
other site with a probability $P(k)$, which decays algebraically
with the distance $k$ as
\begin{equation}
P(k)\sim k^{-\alpha}.
\end{equation}
In such models the infection can be thought of as being spread by
L{\'e}vy flights~\cite{Levy}. Janssen {\textit{et al.}} solved
this problem using field-theoretic renormalization group
calculations~\cite{JOWH99}. The results of this analysis is in
agreement with numerical simulations~\cite{HinrichsenHoward99}.
More recently these studies have been extended to contact
processes with temporal~\cite{Dalmaroni} and spatio-temporal
L{\'e}vy flights~\cite{Adamek05} as well as to systems with
disorder~\cite{TheoGeisel}.

The $\alpha$-process studied in the present work differs from
L{\'e}vy-flight-mediated contact process. Here an inactive site
can only be activated by the nearest active site. Therefore, in
one space dimension, the interaction range is effectively cut off
by half the actual size of the corresponding island of inactive
sites. Thus infections in the $\alpha$-process are mediated by
\textit{truncated} L{\'e}vy flights, which cannot overtake other
active sites. As will be shown below, this truncation changes the
critical properties at the transition significantly. In
particular, although both models exhibit critical exponents which
vary with $\alpha$, the exponents themselves are not the same.
Moreover, while in the L{\'e}vy flights model the exponents become
mean-field like for small $\alpha$ \cite{Hinrichsen00} no such
regime exists in the $\alpha$-process.

In the following section we analyze the $\alpha$-process by a
mean-field approximation as well as numerical simulations and
compare it to contact processes with unrestricted L{\'e}vy
flights. In Sect.~\ref{sec:sigma} we further generalize the model
by taking the rate for nearest-neighbor infection to depend on the
distance to the nearest active site. This rate is controlled by an
additional exponent $\sigma$. The phase diagram of this
generalized model, called $\alpha$-$\sigma$-process, is then
studied by mean-field and numerical methods. Conclusions are
finally presented in Sect.~\ref{sec:conclusions}

\section{Analysis of the model}
\subsection{Mean-field approximation}
\label{subsec:mf}

We now consider the $\alpha$-process within a mean-field
approximation in terms of the density of active sites $\rho(t)$.
To this end the mean-field equation of the ordinary contact
process has to be extended by a term that accounts for long-range
infections according to Eq.~(\ref{process3}). Assuming that the
sites of a one-dimensional lattice are uncorrelated and
independently active with probability $\rho$, it is easy to check
that the probability of an inactive site to lie at a distance $l$
from the nearest active site is $(1-(1-\rho)^2) (1-\rho)^{2 l -
2}$. Summing up the contributions for all distances $l$, the
mean-field equation governing the dynamics of the density of
active sites takes the form
\begin{equation}
\partial_t\rho=-\rho + q \rho (2-\rho) \sum_{l=1}^{\infty}
\frac{(1-\rho)^{2l-1}}{l^{\alpha}} \,.
\label{MFalpha1}
\end{equation}
where the first term appearing on the r.h.s. of Eq.
(\ref{MFalpha1}) corresponds to the short range annihilation
process. Note that the usual DP short range activation term
is given by the first term of the sum in the r.h.s. of
Eq. (\ref{MFalpha1}) \\

Turning the sum into an integral the leading terms in $\rho$ in
the dynamical equation (\ref{MFalpha1}) may be evaluated. It is
found that for $\alpha<1$ and, to leading order in $\rho$, Eq.
(\ref{MFalpha1}) becomes
\begin{equation}
\partial_t\rho=r\rho + u \rho^{\alpha}
\end{equation}
where $(r,u)$ are constants and $u>0$. Since the leading term in
this equation is $\rho^{\alpha}$, and since its coefficient is
positive, the absorbing state $(\rho=0)$ is always unstable and no
transition takes place for any finite $q$. This is also the case
for $\alpha=1$, where logarithmic corrections to the leading
linear terms destabilize the absorbing phase.\\
On the other hand, for $1<\alpha<2$ the leading term in the
equation is the linear one. Moreover the coefficient $u$ of the
leading non-linear term is negative. This results in a continuous
transition to the absorbing state which takes place at $r=0$
corresponding to a non-vanishing rate $q$. Such a phase transition
belongs to a universality class different from the one of the
short range DP model. For example, in the stationary state the
density of active sites scales as
\begin{equation}
\rho_{\rm \scriptscriptstyle stat} \sim r^{1/(\alpha-1)}\,,
\end{equation}
hence the mean-field order parameter exponent, $\beta^{\rm
\scriptscriptstyle MF} = \frac{1}{\alpha-1}$, varies continuously
with $\alpha$. It diverges in the limit $\alpha \to 1$. For
$\alpha\geq2$, however, the leading non-linear term is $ v \rho^2$
where $v$ is a negative constant. The mean-field equation thus
reduces to that of the short range DP process, and usual DP-like
transition is expected.

Thus
\begin{equation}
\beta^{\rm \scriptscriptstyle MF} =
\left\{\begin{array}{ll}
\frac{1}{\alpha-1}
\quad\quad & 1< \alpha < 2 \\
1\quad\quad &\alpha \geq 2 \,.

\end{array}
\right.
\end{equation}

Before discussing the results of the numerical study of this model
we consider a more refined mean-field approximation, where the
effect of the long-range process on the diffusion constant is
taken into account. To proceed we assume that, although the
dynamical process is long range, it does not modify the usual
Laplacian form of the spatial interactions, $D \nabla^2 \rho$. The
reason is that the infection range in the $\alpha$--model is
cut-off by the maximal distance to the nearest active site, making
it effectively of finite range. However, as we shall see below,
the effect of the $\alpha$ infection process is to make the
diffusion coefficient $D$ dependent on the density $\rho$.

In order to evaluate $D(\rho)$ we note that the infection process
is cut off at length-scales $l_{\rm av}\sim1/\rho$, which is the
average distance between active particles in a system with density
$\rho$. The mean square displacement of an infection is expected
to scale as
\begin{equation}
\langle l^2 \rangle \sim \int_{\Lambda}^{l_{\rm av}} dl\,
l^{2-\alpha} \sim \rho^{\alpha-3} + D_0. \label{l2}
\end{equation}
where $\Lambda$ is a cutoff due to lattice spacing from which the
usual constant contribution $D_0$ (positive for $\alpha > 3$) to
the diffusion coefficient arises. Since $D(\rho)\sim \langle l^2
\rangle$, Eq. (\ref{l2}) suggests that the effective diffusion
constant diverges for $\alpha<3$. Thus the space-time-dependent
version of the mean-field equation for $1 < \alpha < 2$ reads to
lowest order
\begin{equation}
\label{MFimproved}
\partial_t\rho(\xvec,t)=r\rho(\xvec,t) + u[\rho(\xvec,t)]^\alpha + [\rho(\xvec,t)]^{\alpha-3}\nabla^2\rho(\xvec,t).
\end{equation}
For this equation, dimensional analysis yields the complete set of
mean-field exponents
\begin{equation}
\nu_\perp^{\rm \scriptscriptstyle MF} = \beta^{\rm \scriptscriptstyle MF}= \frac{1}{\alpha-1}\,, \quad \nu_\parallel^{\rm \scriptscriptstyle MF}=1\,,
\end{equation}
where as usual $\nu_\perp^{\rm \scriptscriptstyle MF}$ and
$\nu_\parallel^{\rm \scriptscriptstyle MF}$ are the mean-field
values for the spatial and temporal exponents which control the
power-law divergence of correlations (respectively in space and
time) at criticality.

On the other hand for $2<\alpha<3$ the non-linear term in the
dynamical equation $u\rho^\alpha$ becomes $v\rho^2$ while the
diffusion coefficient remains singular at $\rho=0$. This implies
that the DP regime sets up only for $\alpha\geq 3$, and not for
$\alpha>2$ as suggested by the analysis of the reaction terms
alone. In particular, for $2<\alpha\leq 3$ lowest order mean-field
reads
\begin{equation}
\label{MFimproved2}
\partial_t\rho(\xvec,t)=r\rho(\xvec,t) + v[\rho(\xvec,t)]^2 + [\rho(\xvec,t)]^{\alpha-3}\nabla^2\rho(\xvec,t).
\end{equation}
and by dimensional analysis one gets
\begin{equation}
\nu_\perp^{\rm \scriptscriptstyle MF} = 2 - \frac{\alpha}{2}\,, \quad \beta^{\rm \scriptscriptstyle MF}= \nu_\parallel^{\rm \scriptscriptstyle MF}=1\,.
\end{equation}
For the dynamical exponent $z=\nu_\parallel / \nu_\perp$, we
obtain
\begin{equation}
\label{zMF}
z^{\rm \scriptscriptstyle MF}= \left\{
\begin{array}{ll}
\alpha -1  \quad\quad & 1< \alpha < 2 \\
 \frac{2}{4-\alpha} \quad\quad& 2 \leq \alpha \leq 3 \\
2 \quad\quad& \alpha > 3 \\
\end{array}
\right.
\end{equation}
In the following we compare these values with the results of extensive
numerical simulations.

%
\subsection{Numerical results}
\label{subsec:alphanumerics}

In order to determine the critical exponents numerically, we
performed Monte-Carlo simulations of the one-dimensional
$\alpha$-process with periodic boundary conditions, starting with
a fully occupied lattice. The time step $dt$ of a single move in
the random sequential updating procedure has been
taken as $1/L$, where $L$ is the lattice size. \\
For models with long-range interactions such simulations are
difficult to perform because of strong finite-size effects, which
increase as  $\alpha \to 1$. In the present model these
finite-size effects manifest themselves by sudden transitions into
the absorbing state. This leads to uncontrollable fluctuations in
the statistical averages. To circumvent this problem we combined
finite size scaling analysis with numerical simulations. The
simulations were carried out on large systems with homogeneous
initial states approaching the critical point from the active phase. \\
The critical point $q_c$ and the dynamical exponent $z$ have been
determined by finite size analysis of the average absorbing time
$\tau$ for an initially fully occupied lattice. At the
critical point $\tau$ is expected to scale with the system size $L$ as
\begin{equation}
\tau \sim L^z.
\label{tauz}
\end{equation}
For one dimensional systems belonging to the DP universality class
the dynamical exponent is known to be $z^{\scriptscriptstyle\rm
DP}=1.580745(10)$ \cite{Jensen99}. We also measured the exponent
$\delta = \beta/\nu_\parallel$, which is associated with the
temporal decay of the density $\rho$ of active sites at
criticality. This calculation has been performed by applying the
method of finite--size analysis proposed by de Oliveira and
Dickman~\cite{Dickman}: whenever the system falls into the
absorbing state the dynamical process is recovered by
automatically resetting the system into a ``typical'' active
configuration. In practice, this procedure suppresses the large
fluctuations which result from the abrupt transitions to the
absorbing state in finite systems. A simple scaling argument shows
that the stationary density at criticality should scale with the
system size as
\begin{equation}
\rho_{stat} \sim L^{-\delta z}.
\end{equation}
Finite size analysis has been performed by averaging over
$10^3$-$10^4$ independent runs.

\begin{figure}
\centerline{\includegraphics[width=114mm]{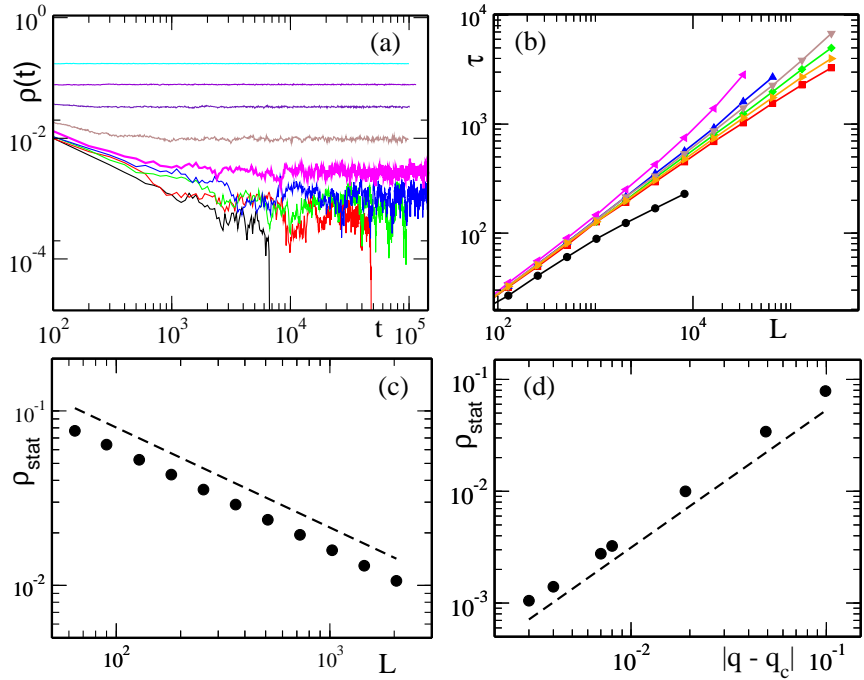}}
\caption{\label{fig:example} \small Results of the numerical
analysis of the $\alpha$-process for $\alpha=1.5$. All plots are
in double-logarithmic scale. a) Density of active sites $\rho(t)$
as a function of time measured for a system of size $L=2^{20}$
starting from a fully occupied lattice. In the active phase and close to the critical value
$q_c=0.4695(5)$ (from top to bottom $q=0.679,
\;0.569,\;0.519,\;0.489,\;0.477,\;0.473,\;0.472,\;0.471,\;0.470$)
finite-size fluctuations yield sudden transitions to the absorbing
state. b) Finite-size scaling analysis of the average absorbing
time $\tau$. From top to bottom
$q=0.475,\;0.471,\;0.470,\;0.469,\;0.468,\;0.467,\;0.45$. c)
Finite-size scaling analysis of the stationary density
$\rho_{stat}$ (see text). d) Stationary density $\rho_{stat}$ as a
function of the distance from $q_c$~. The dashed lines in panels
c-d are the best fit (see table \ref{tab:alphaexp}).}
\end{figure}

For $\alpha > 2.0$ the estimates of the exponent $\delta$,
obtained by finite size analysis, have been compared with the
standard  analysis, based on the time-decay of $\rho(t)$ at
criticality. Usually, large fluctuations make the latter method
quite inaccurate and time consuming. In order to obtain better
performances we carried out numerical simulations of large systems
($L=2^{20} \sim 2^{21}$) and over long time lapses ($10^5 \sim
10^6$ time steps), while averaging over a few different
realizations. Good agreement with the predictions of finite--size
analysis is found.

 The critical exponent $\beta$ has been
determined by measuring the stationary density of active sites
$\rho \sim (q-q_c)^\beta$ in large systems ($L=2^{15} \sim
2^{18}$) for different values of $q$, just above $q_c$.

The estimated values for the exponents $\delta$ and $\beta$ can be
compared with the best numerical estimates of the DP exponents in
one dimension, $\delta^{\scriptscriptstyle\rm DP}=0.159464(6)$ and
$\beta^{\scriptscriptstyle\rm DP}=0.276486(8)$ \cite{Jensen99}.
Typical curves resulting from this numerical procedure are shown
in Fig. \ref{fig:example} for $\alpha=1.5$.\\

\begin{table}
\begin{center}
\begin{tabular}{|l|l|l|l|l|l|l|}
$\alpha$ &  $q_c$ &     $z$ &       $\delta$ &  $\beta$
& $\beta^{\rm \scriptscriptstyle MF} = \delta^{\rm \scriptscriptstyle MF}$
& $z^{\rm \scriptscriptstyle MF}$ \\ \hline
$1.2$ &     $0.205(3)$ &  $0.34(2)$ &  $2.0(1)$ &   $2.73(15)$ & $5$ & $0.2$\\
$1.5$ &     $0.4695(5)$ & $0.67(2)$ &  $0.84(4)$ &  $1.25(5)$ & $2$ & $0.5$\\
$1.8$ &     $0.714(1)$ &  $0.99(5)$ &  $0.46(3)$ &  $0.68(7)$ & $1.25$ &$0.8$\\
$2.0$ &     $0.8592(2)$ & $1.21(2)$ &  $0.31(1)$ &  $0.49(2)$ & $1$ & $1$\\
$2.1$ &     $0.9250(3)$ & $1.25(2)$ &  $0.28(1)$ &  $0.46(2)$ & $1$ & $1.0526\ldots$\\
$2.3$ &     $1.0453(2)$ & $1.43(2)$ &  $0.22(1)$ &  $0.36(1)$  &$1$ & $1.1764\ldots$\\
$2.5$ &     $1.1492(2)$ & $1.48(4)$ &  $0.19(1)$ &  $0.32(1)$  & $1$ & $4/3$\\
$2.6$ &     $1.1955(2)$ & $1.54(3)$ &  $0.175(10)$ &  $0.30(1)$  & $1$ & $1.4285\ldots$\\
$2.7$ &     $1.2381(2)$ &  $1.56(2)$ & $0.17(1)$ &  $0.294(8)$  & $1$ & $1.5384\ldots$\\
$3.0$ &     $1.3470(2)$ &  $1.58(2)$ & $0.166(8)$ & $0.278(8)$  & $1$ & $2$
\end{tabular}
\end{center}
\caption{\label{tab:alphaexp}
Estimates of the critical points and exponents for various values of $\alpha$. 
These values are plotted in Fig.~\ref{fig:alphaexp}. For comparisons mean field 
values are also reported.}
\end{table}

\begin{figure}
\centerline{\includegraphics[width=114mm]{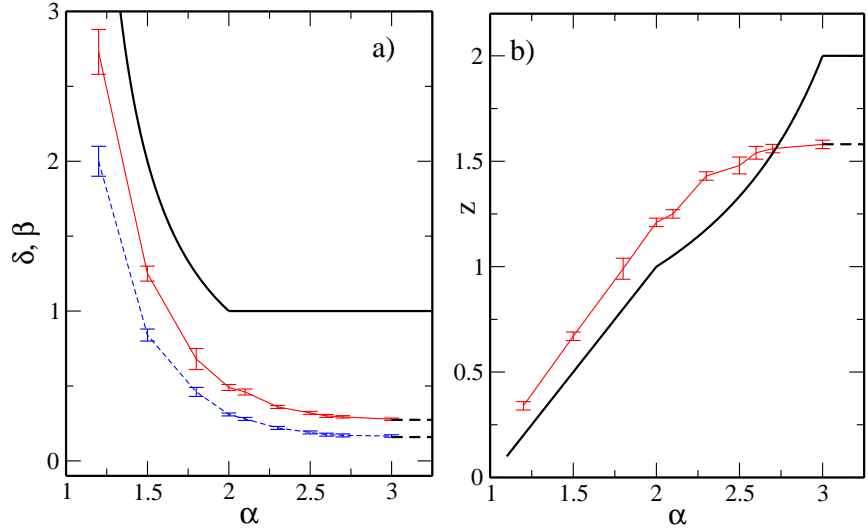}}
\caption{\label{fig:alphaexp} \small Comparison between numerical
estimates (including error bars) and the mean-field predictions
for the critical exponents as a function of the control parameter
$\alpha$. a) The exponents $\beta$ (thin solid line),
$\delta=\beta/\nu_\parallel$ (thin dashed line) and the mean-field
prediction (thick solid line); note that
$\beta^{\scriptscriptstyle\rm MF} = \delta^{\scriptscriptstyle\rm
MF}$. b) The dynamical exponent $z=\nu_\parallel/\nu_\perp$ (thin
solid line) and the mean-field prediction (thick solid line). The
thick dashed lines on the right of both plots mark the best
numerical estimates of the DP exponents in 1D.}
\end{figure}

As summarized in Table~\ref{tab:alphaexp}, the numerical results
show a qualitative agreement with the mean-field predictions.
In particular, it is found that all exponents
vary as $\alpha$ is increased up to a critical value $\bar{\alpha}$,
beyond which they become consistent (within numerical error)
with the DP values. As $\alpha$
approaches $1$, the exponents $\beta$ and $\delta$ are found to
increase, while the dynamical exponent $z$ vanishes as $\alpha -1$.
The comparison between numerics and mean-field
is shown in Fig.~\ref{fig:alphaexp}.

Note that according to numerical results the crossover to the DP
scaling regime seems to take place at $\bar{\alpha}\approx 2.7$.
On the other hand, the mean-field predicts that the crossover
takes place in two steps: first, $\beta$ and $\delta$ become
DP-like at $\alpha =2$ and then the exponent $z$ assumes its DP
value at $\alpha =3$~.

\subsection{Comparison with unrestricted L{\'e}vy flights}
\label{subsec:levy}

As anticipated in the Introduction, it is instructive to compare
the results of the $\alpha$-process with those of contact
processes with infections mediated by \textit{unrestricted}
L{\'e}vy flights (see Refs.~\cite{JOWH99,HinrichsenHoward99}). In
the latter case three different dynamical regimes were identified:
\begin{itemize}
\item a mean-field regime for $\alpha < 3d/2$ characterized by the
critical exponents $\beta^\MF=1$, $\nu_\perp^\MF=(\alpha-d)^{-1}$,
and $\nu_\parallel^\MF=1$, where $d$ denotes the spatial
dimension. \item a non-trivial phase for $3d/2 \leq \alpha \leq
\alpha^*$ with continuously varying exponents, restricted by the
additional scaling relation
$\nu_\parallel-\nu_\perp(\alpha-2d)-2\beta=0$. \item a DP regime
for $\alpha>\alpha^*$.
\end{itemize}
Here the upper threshold for $\alpha$ can be expressed in terms of
DP exponents by $\alpha^*=2d+z^\DP-2\beta^\DP/\nu_\perp^\DP\approx
3.0766(2)$. The phase structure of this model exhibits a mean
field regime for $\alpha<3d/2$ which is not present in the
$\alpha$-process. In this mean field regime the unrestricted
L{\'e}vy flights become so long-ranged that spatial fluctuations
are destroyed, leading to a homogeneous distribution of active
sites. In the present model, however, the long-range interactions
are effectively cut off at the distance to the nearest active
site, hence there is no direct communication by long-range
interactions between adjacent intervals of inactive sites. This
restriction yields non-trivial correlations.

It is instructive to compare mean-field
equation~(\ref{MFimproved}) with the corresponding equation for
unrestricted L{\'e}vy flights, which -- according to the present notation
-- can be written as
\begin{equation}
\label{MFLevy}
\partial_t\rho(\xvec,t)=n\rho(\xvec,t) - \rho^2(\xvec,t) + D\nabla^{\alpha-1}\rho(\xvec,t).
\end{equation}
Here $\nabla^{\alpha-1}$ is a symetric fractional derivative, which
generates unrestricted L{\'e}vy flights. Clearly, this equation
has scaling properties different from (\ref{MFimproved})~. For instance,
within the mean-field
approximation the exponent $\beta$ is equal to $1$. Therefore, it
is likely(?) that also in low dimensions models with truncated and unrestricted
L{\'e}vy flights belong to different universality classes.

\subsection{The $\alpha$-process in higher dimensions}
\label{sec:high-dim}

The $\alpha$-process and its mean-field equation can be easily
generalized to $d>1$. In this case the range, $l$, of the
activation process of an inactive site located at $i$ is given by
the radius of the maximal "empty" (inactive) sphere centered in
$i$. Within the mean-field approximation the average volume of the
"empty" spheres is proportional to $1/\rho$, so that the average
radius of empty spheres is $<l> \propto \rho^{-1/d}$. Moreover,
the diffusion coefficient $D(\rho)$ can be easily computed by
extending Eq. (\ref{l2}) to arbitrary dimension $d$
\begin{equation}
D(\rho) \sim \rho^{\frac{\alpha-2}{d} -1} + D_0
\end{equation}
Therefore, the corresponding mean-field equation reads
\begin{equation}
\label{MFAlpha3}
\partial_t\rho=r\rho - u\rho^{\alpha/d} -v \rho^2 +
D(\rho) \nabla^2 \rho + o(\rho^2)\,,
\end{equation}
where $r(q)$ vanishes at the critical point while $u(q)$ and
$v(q)$ are positive constants. As in the 1$d$ case, if $\alpha<d$
the absorbing state cannot be reached and there is no transition.
On the other hand, for $\alpha > 2d$  and $d \geq 2$ a DP regime
sets in, where both leading terms in the dynamical equation and in
the diffusion constant become DP-like. In the range $d<\alpha<2d$
the system displays a non-trivial transition, with continuously
varying
critical exponents. \\
In summary, the mean-field exponents for $d \geq 2$ are
\begin{equation}
\nu_\parallel^{\rm \scriptscriptstyle MF} = 1\,,
\end{equation}
\begin{equation}
\beta^{\rm \scriptscriptstyle MF} =
\delta^{\rm \scriptscriptstyle MF} \nu_\parallel^{\rm \scriptscriptstyle MF}
= \left\{
\begin{array}{ll}
\frac{1}{\alpha/d - 1} \quad\quad & d< \alpha < 2d \\
1 \quad\quad& \alpha > 2d
\end{array}
\right.
\end{equation}
and
\begin{equation}
\nu_\perp^{\rm \scriptscriptstyle MF}
= \frac{\nu_\parallel^{\rm \scriptscriptstyle MF}}{z^{\rm \scriptscriptstyle MF}} = \left\{
\begin{array}{ll}
\frac{1}{\alpha - d} \quad\quad & d< \alpha < d+2 \\
 \frac{1}{2} \quad\quad& \alpha \ge d+2
\end{array}
\right.
\end{equation}

\section{Generalization of the model}
\label{sec:sigma}

\subsection{The $\sigma$-process}
\label{subsec:sigma}

In a recent paper we studied a different version of a contact
process with long-range interactions, called
$\sigma$-process~\cite{sigma}. In this model infections are
short-ranged, i.e., active sites can only activate their nearest
neighbors, but the \textit{rate} for short-range infections
depends algebraically on the distance $m$ to the nearest active
site before the update. Specifically, the $\sigma$-process is
defined by the transition rates
\begin{eqnarray}
\label{sigma1}
1  &\rightarrow& 0      \hspace{8mm}\mbox{with rate }\hspace{0.5cm} 1 \\
\label{sigma2}
10,01 &\rightarrow& 11  \hspace{5mm}\mbox{with rate }\hspace{0.5cm} q(1+a/m^\sigma)
\end{eqnarray}
where $m$ denotes the distance to the next active site. Here $a$
is a constant, $q$ is again the control parameter, and $\sigma$ is
an exponent controlling the characteristics of the interaction. It
turns out that for $\sigma>1$ the transition belongs to the DP
universality class, while for $0<\sigma<1$ the transition becomes
first order.

The model was introduced as a toy model in order to explain why
non-equilibrium wetting processes, with a sufficiently strong
attractive short-range force between interface and substrate, can
exhibit a first-order transition. In this case the growth rate in
the bulk is positive, so that spontaneous pinning far away from
the edges can be neglected, meaning that the transition is driven
by unpinning and spontaneous growth of bound regions. The binding
rate, however, was found to vary with the actual size of the
unpinned regions, which motivates the algebraic $m$-dependent rate
in Eq.~(\ref{sigma2}). It should be noted that the
$\sigma$-process is mainly relevant for non-equilibrium wetting in
one spatial dimension only, where pinned sites effectively
separate depinned islands into non communicating regions.
Moreover, there is no unique and straightforward way to extend
such a process in higher dimensions. In the following we combine
the two long-range processes in a single $1d$ model and study the
corresponding phase diagram in the $\alpha$-$\sigma$-plane.

\subsection{Definition of the $\alpha$-$\sigma$-process}
\label{subsec:alphasigma}

The combined $\alpha-\sigma$ model, which evolves  by random
sequential update over a time step $dt = 1/L$ is defined by the
following transition rates:
\begin{eqnarray}
\label{combprocess1}
1  &\rightarrow& 0      \hspace{8mm}\mbox{with rate }\hspace{0.5cm} 1 \\
\label{combprocess2}
10,01 &\rightarrow& 11  \hspace{5mm}\mbox{with rate }\hspace{0.5cm} q(1+a/m^\sigma)\\
\label{combprocess3}
000 &\rightarrow& 010   \hspace{7mm}\mbox{with rate }\hspace{0.5cm} bq/l^\alpha
\end{eqnarray}
where $m$ is the size of the inactive island and $l$ denotes the
distance from the nearest active site. Here $q$ is the control
parameter, $\alpha$ and $\sigma$ are the control exponents, and
$a$ and $b$ are constants. The ``pure'' $\alpha$- and $\sigma$-
processes are recovered in the limits $\sigma \to \infty$ and
$\alpha \to \infty$, respectively.

\subsection{Mean-field analysis and numerical simulations}
\label{subsec:asmf}
Neglecting spatial fluctuations, the mean field equation of the
combined $\alpha - \sigma$-process (see Sec.~\ref{subsec:mf} and
~\cite{sigma}) becomes
\begin{equation}
\partial_t\rho \;=\;
-\rho \,+\,
q \rho^2 \sum_{m=1}^{\infty} (1+\frac{a}{m^{\sigma}})(1-\rho)^m  \,+\,
b q \rho (2 - \rho) \sum_{l=2}^{\infty} \frac{(1-\rho)^{2l-1}}{l^{\alpha}} \,.
\label{MFsigmaalpha}
\end{equation}
In the regime of interest, namely for $0<\sigma<1$ and
$1<\alpha<2$, the mean-field equation can be written to leading
order as
\begin{equation}
\label{MFSigmaAlpha}
\partial_t\rho=r\rho + p\rho^{1+\sigma} - u\rho^\alpha + 0(\rho^2)\,,
\end{equation}
where $r,p$ and $u$ are $q$-dependent constants:
\begin{eqnarray}
r &=& q-1+bq{2^{2-\alpha}\over{\alpha-1}}\,, \\
p &=& aq\,\Gamma(1-\sigma)\,,\\
u &=& bq\,{{2^{\alpha-1}}\over{\alpha-1}}\Gamma(2-\alpha).
\end{eqnarray}
%
%

Note that these are approximate expressions, as they are obtained
by substituting the sums in (\ref{MFsigmaalpha}) by integrals to
obtain simple analytic expressions. This approximation does not
affect the sign of $p$ and $u$ which turn out to be positive
constants in the regime of interest. Accordingly, if the  leading
nonlinear term is $\rho^{1+\sigma}$ (i.e. $\sigma<\alpha-1$)~, the
transition is first order. On the other hand, if the leading
nonlinear term is $\rho^{\alpha}$ ($\sigma>\alpha-1$)~, the
critical behavior of the pure $\alpha$-model is recovered. Thus,
the mean-field approach predicts that the critical exponents are
independent of $\sigma$~. One can add the effect of the long range
processes on the diffusion constant as was done in Sec.
\ref{subsec:mf}. It is readily seen that the critical exponent $z$
assumes its DP value for $\alpha>3$ as long as $\sigma>1$.

\begin{figure}
\centerline{\includegraphics[width=114mm]{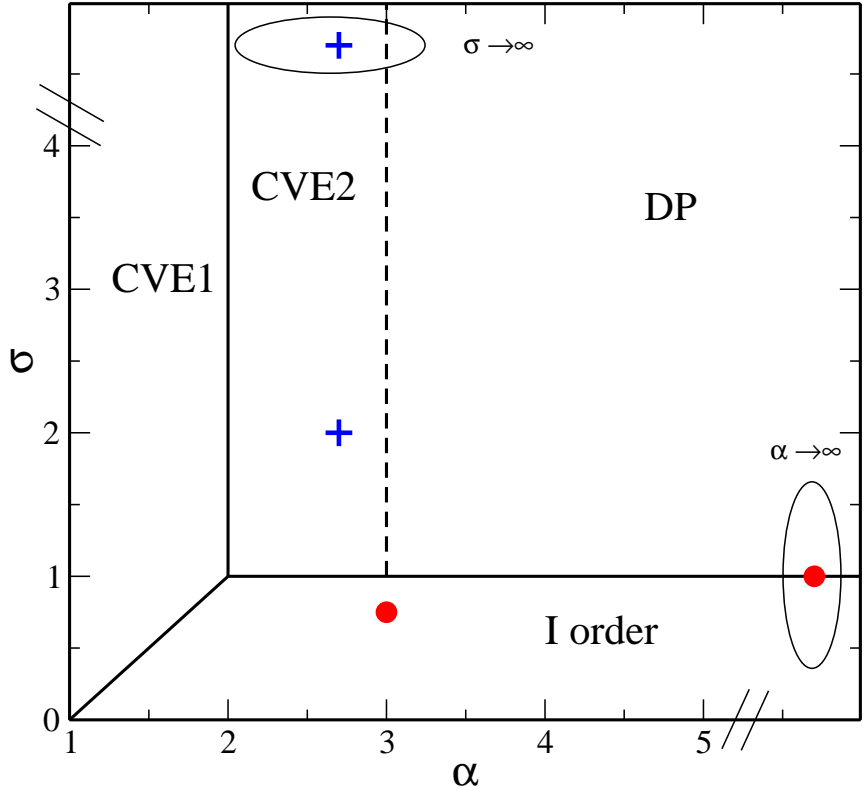}}
\caption{\label{fig:phasediag} \small Mean-field phase diagram of
the $\alpha$-$\sigma$-process. Four different regions are
identified corresponding to different types of transitions to the
absorbing state: a DP region, a region of a first order transition
and two regions of continuously varying critical exponents as
explained in the text. We also indicate the of the phase boundary
lines as determined by numerical simulations: full dots for the
DP-first order boundary and crosses for the DP-CVE boundary. The
numerical results suggest the existence of a single CVE phase with
a single DP-CVE line located at $\bar \alpha \approx 2.7$~.}
\end{figure}

The mean-field phase diagram is shown in Fig.~\ref{fig:phasediag}.
It contains four different phases:
\begin{enumerate}
\item [(i)]a DP phase for $\sigma>1,\alpha>3$,

\item [(ii)] a continuously varying exponent phase CVE1 for $\sigma> \alpha -1$ and $1<\alpha <
2$ where the transition is second order with continuously varying
exponents $\beta, \delta$ and $z$.

\item [(iii)] a continuously varying exponent phase CVE2 for $\sigma > 1$ and $2< \alpha < 3$
where the transition is second order but where $z$ is the only
continuously varying exponent.

\item [(iv)]a phase for $\sigma < \alpha - 1$ and $0< \sigma < 1$

where the transition is discontinuous.
\end{enumerate}

\begin{table}
\begin{center}
\begin{tabular}{c|cccccc}
$\alpha$ & 2.0 & 2.3 & 2.5 & 2.6 & 2.7 & 3.0 \\
\hline
$q_c$ & $0.7783(2)\quad$ & $0.9308(3)\quad$ & $1.0155(3)\quad$ &
$1.0530(3)\quad$ & $1.0880(5)\quad$ & $1.1785(5)\quad$\\
\hline
$z$ & 1.18(3) & 1.40(3) & 1.51(3) & 1.50(4) & 1.54(5) & 1.57(3) \\
\hline
$\delta$ & 0.32(2) & 0.22(1) & 0.19(1) & 1.18(1) & 1.18(1) & 1.16(1) \\
\hline
$\beta$ & 0.52(1) & 0.37(3) & 0.32(3) & 0.30(2) & 0.29(1) & 0.28(2) 
\end{tabular}
\end{center}
\caption{\label{tab:alphasigmaexp} $\alpha-\sigma$ process: 
Estimates of the critical points and exponents for $\sigma=2$ and various values
 of $\alpha$.
Other parameters have been fixed to $a=2$ and $b=1$}
\end{table}

Numerical simulations have been performed for testing the
predictions of the mean-field phase diagram. In these simulations
we have taken $a=2$ and $b=1$~.


An extensive numerical study of the phase-diagram of the
$\alpha$-$\sigma$-process is time consuming and was not tackled in
the present paper. Rather, we restricted our analysis to a small
number of cuts in the $\alpha$-$\sigma$ plane in order to test the
validity of the mean-field predictions. We find that while the
general features of the phase-diagram are correctly reproduced by
the mean-field, the location of the transition lines and the
values of the critical exponents are changed.

The numerical method proposed in ~\cite{Dickman} has been applied
for the inspection of the phase diagram along the line $\sigma =2$
for a discrete set of values of $\alpha$ ranging between 2 and 3~.
In these simulations we have used the same system sizes and
conditions described in ~\ref{tab:alphasigmaexp}. The numerical
simulations suggest that in fact there is a single phase with
continuously varying exponents (CVE) rather than two as predicted
by the mean field. In this phase all critical exponents are
continuously varying with $\alpha$. Both in the DP and CVE regions
the critical exponents coincide, within numerical accuracy, with
those found for the pure $\alpha$-model (i.e., $\sigma \to
\infty$). The results reported in Table~\ref{tab:alphasigmaexp}
are in agreement with the mean-field prediction that scaling
properties in the critical regions are independent of the
$\sigma$-process. It is worth stressing that even for finite
values of $\sigma$ the boundary between the DP and the CVE region
is located close to $\bar\alpha \approx 2.7$~. The position of the
boundary line for $\sigma=\infty$ and $\sigma=2$ is marked by
crosses on the phase diagram (see Fig. \ref{fig:phasediag})

We now consider the boundary between the first--order and the DP
regions. For the pure $\sigma$-model (i.e., $\alpha \to \infty$)~
this boundary was investigated numerically in Ref. \cite{sigma}~.
Following (\cite{sigma}) we determine the nature of the phase
transition for finite $\alpha$ by studying the size distribution
of inactive domains in stationary active state. At the transition
to the absorbing state this distribution is expected to exhibit a
power law tail $P(m) \sim m^{-\gamma}$ for large inactive domains
of size $m$. By numerically determining the exponent $\gamma$, the
nature of the transition may be deduced. For $\gamma > 2$ the
average domain size is finite at the transition and thus it is
first order. On the other hand for $\gamma \le 2$ the average
domain size diverges at the transition and thus the transition is
continuous.  In particular, for DP, where the transition is
continuous, one has $\gamma^{\rm
\scriptscriptstyle DP}=2-\beta/\nu_\perp \simeq 1.747$~. \\

We have analyzed the nature of the phase transition for $\alpha =
3$ at some values of $\sigma$ ranging between 0.5 and 2~. We find
that $\gamma = \gamma^{DP}$ for $\sigma > 0.8$ indicating that the
transition to the absorbing state is of DP nature. On the other
hand for $\sigma < 0.8$ we find $\gamma > 2$, indicating a first
order transition. The position of the boundary line for
$\alpha=\infty$ and $\alpha=3$ is marked by dots on the phase
diagram (see Fig. \ref{fig:phasediag}).

\begin{figure}
\centerline{\includegraphics[width=114mm]{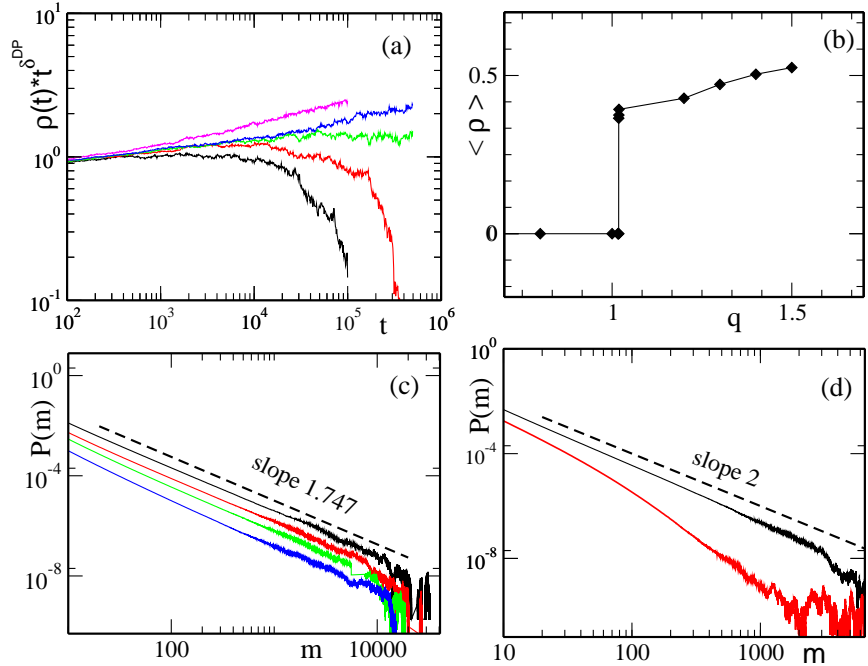}}
\caption{\label{fig:sigma} \small Numerical study of the
$\alpha$-$\sigma$-process for $\alpha=3.0$. a) $\rho(t)
t^{\delta^{DP}}$ vs $t$ for $\sigma=0.8$ and values of the control
parameter close to $q_c \approx 1.043$~: from top to bottom
$q=1.045,\;1.044,\;1.043,\;1.042,\;1040$~. b) stationary density
$\rho_{stat}$ as a function of  $q$ for $\sigma=0.7$: a jump
occurs at the first-order transition point $q_c \approx 1.0184$~.
In c) and d) we display the probability distribution of the size
of inactive domains $P(m)$ at the transition. c) from top to
bottom $\sigma = 1.5,\, 1.1,\, 0.9,\, 0.8)$~: the power-law decay
agrees with the DP scaling (see dashed line). d) from top to
bottom $\sigma = 0.7,\, 0.5$~: $P(m)$ decays faster then $m^{-2}$
so that $\langle m \rangle$ is finite. In both of these plots the
curves are not normalized and have been rescaled to fit on the
same plot.}
\end{figure}

Details of the numerical analysis is presented in
Fig.~\ref{fig:sigma}. In this figure we first illustrate the
method for identifying the transition point. This is done by
plotting $\rho(t)t^{\delta_{DP}}$
 as a function of $t$ for
different values of $q$ and searching for the $q$ value for which
this quantity approaches a constant. The data were obtained for
for quite large system sizes ($L \approx 2^{18} \sim 2^{19}$) and
homogeneous initial conditions. We then display the distribution
function $P(m)$ at the transition, and determine the exponent
$\gamma$ which controls its large $m$ behavior. The value of the
control parameter $q$ at the transition point is given in
Table~\ref{tab:alphasigmacp} for $\alpha=3$ and several values of
the parameter $\sigma$.

Our numerical results suggest that the boundary between first
order and DP like behavior varies with $\alpha$. In particular it
takes place at $\sigma \simeq 0.8$ for $\alpha = 3$ while it is
$\sigma \simeq 1$ in the limit of infinite $\alpha$. Thus the
boundary line in the $\alpha - \sigma$ plane is not parallel to
the $\alpha$ axis at variance with mean field prediction.


\begin{table}[tbc]
\begin{center}
\begin{tabular}{c|cccccc}
$\sigma$ & 0.5 & 0.7 & 0.8 & 0.9 & 1.1 & 1.5 \\
\hline
$q_c$ & $0.9515(5)\quad$ & $1.0184(1)\quad$ & $1.0430(5)\quad$ & $1.0647(3)\quad$ & $1.0992(1)\quad$ & $1.1445(5)\quad$
\end{tabular}
\end{center}
\caption{\label{tab:alphasigmacp} $\alpha-\sigma$-process:
estimates of the critical points $q_c$ for $\alpha=3$, $a=2$,
$b=1$ and various values of $\sigma$.}
\end{table}
\section{Conclusions}
\label{sec:conclusions}
In many cases it is useful to project complex dynamical processes
onto simpler ones which could be more readily analyzed. In doing
so it may happen that the local dynamics of the original model is
translated into a non-local dynamics of the projected model. For
example it has been argued that when the dynamics of fluctuating
interfaces interacting with a wall is projected onto the dynamics
of a contact process different types of long range processes may
take place.

Motivated by this observation we introduced a non-local version of
the contact process where particles are created with a rate which
decays like $l^{-\alpha}$ with the distance $l$ from the nearest
particle. Mean-field analysis shows that the critical exponent
$\beta$ associated with the transition to the absorbing state
varies continuously with $\alpha$ for $1<\alpha<2$ and it diverges
for $\alpha$ approaching $1$. The usual short range mean-field
exponent $\beta=1$ is recovered above the threshold $\alpha=2$.
The exponent $\nu_\parallel$ is found to be $1$ in the entire
range. Numerical studies in one dimension support this general
behavior, although the value of $\beta$ and the upper threshold
seem to be different.

By including spatial fluctuations in the mean-field equations the
mean-field exponents $\nu_\perp$ and $z$ are evaluated. It is
found that near $\alpha=1$ the exponent $z$ vanishes linearly,
while it approaches the usual mean-field value $z=2$ at another
threshold, which in one dimension is $\alpha=3$. Also in this case
we find a qualitative agreement with numerics.\\
Extension of the model to include another long range dynamical
process, the $\sigma$-process, has also been considered using both
mean-field approximation and numerical simulations. Here, too, the
global features of the mean-field phase diagram are qualitatively
recovered by the numerical study.

Acknowledgements

We all thank A. Politi for many fruitful discussions. The support of
the Israel Science Foundation (ISF) and the Einstein Center for
Theoretical Physics is gratefully acknowledged. Four of us
(FG,HH,RL and DM) would like to thank the Newton Institute in
Cambridge (UK) for the kind hospitality during the programme
"Principles of the Dynamics of Non-Equilibrium Systems".

\end{document}